\documentclass{article}

\usepackage{graphicx}
\usepackage{amssymb}
\usepackage[latin1]{inputenc}

\begin{document}




\begin{center}

\LARGE
Fine frequency shift of single vortex entrance and exit in superconducting loops\\

\vspace{1.2cm}

\large

Florian R. Ong$^a$\footnote{email: florian.ong@grenoble.cnrs.fr}, 
Olivier Bourgeois$^a$\footnote{email: olivier.bourgeois@grenoble.cnrs.fr},
Sergey E. Skipetrov$^b$,
Jacques Chaussy$^a$,
Simona Popa$^c$,
Jérôme Mars$^c$,
Jean-Louis Lacoume$^c$

\normalsize

\vspace{1cm}

\begin{itemize}
\it

\item[a] Institut N\'{e}el, CNRS, Laboratoire associé à l'Universit\'{e} Joseph Fourier, 38042 Grenoble, France

\item[b] Laboratoire de Physique et Mod\'elisation des Milieux Condens\'es,
Maison des Magist\`{e}res, CNRS and Universit\'{e} Joseph Fourier, 38042
Grenoble, France

\item[c] Laboratoire des Images et des Signaux, CNRS, Institut National Polytechnique de Grenoble and Universit\'{e} Joseph Fourier, Domaine Universitaire, 38402 Saint Martin d'H\`eres, France

\end{itemize}

\end{center}


\begin{abstract}

\normalsize

The heat capacity $C_{p}$ of an array of independent aluminum rings has been measured under an external magnetic field $\vec{H}$ using highly sensitive ac-calorimetry based on a silicon membrane sensor. Each superconducting vortex entrance induces a phase transition and a heat capacity jump and hence $C_{p}$ oscillates with $\vec{H}$. This oscillatory and non-stationary behaviour measured versus the magnetic field has been studied using the Wigner-Ville distribution (a time-frequency representation). It is found that the periodicity of the heat capacity oscillations varies significantly with the magnetic field; the evolution of the period also depends on the sweeping direction of the field. This can be attributed to a different behavior between expulsion and penetration of vortices into the rings. A variation of more than 15\% of the periodicity of the heat capacity jumps is observed as the magnetic field is varied. A description of this phenomenon is given using an analytical solution of the Ginzburg-Landau equations of superconductivity.

\end{abstract}

\vspace{0.4cm}


\vspace{0.4cm} 

\vspace{0.4cm} 
doi : 10.1016/j.physc.2007.05.050

\newpage


\normalsize

\section{Introduction}

Thanks to recent evolutions in micro and nanofabrication processes, new devices are emerging which permit the study of the specific heat of mesoscopic systems even for samples having very small mass (typically under 100 nanogram)\cite{riou97,fom97,lindell,bourgprl,ongPRB2006,rouknano}. This kind of innovative sensor opens up a great possibility of studies which can be carried out on nanosystems. It enables the study of the phase diagram and the phase transitions specific to systems of small size. Indeed, the reduction of the dimensions of superconducting or magnetic systems leads to the appearance of new phase transitions \cite{buzdin95}, or mechanisms proper to quantum physics which can be studied from a thermal point of view \cite{fink81,buzdin2005,valls2005}. More specifically, the study of vortex matter in nano-engineered arrays of mesoscopic superconducting systems is of particular interest. The effect of the geometry, the thickness of the film or the size of the system compared to a physical characteristic length scale can be studied from a thermal point of view in order to better understand the physics of vortex in extreme or particular conditions. In general calorimetry has the advantage of being sensitive to any energy level in the system. Magnetization for instance is sensitive only to the magnetic aspect of the sample. Through calorimetric measurements any phase transition whatever its origin is expected to exhibit a signature.

In a superconducting ring a supercurrent appears to screen or enhance a perpendicular external magnetic field (described by a vector potential $\vec{A}$), so as the fluxoid $\phi'$ is fixed to an integer multiple of $\phi_0=h/2e$ (the superconducting quantum flux); $\phi'$ is then given by:

\begin{equation}
\phi' = \oint \left[ \frac{m^*}{e^*}\vec{v_s} + \vec{A} \right]\cdot d\vec{l}
 = \oint \left[ \frac{m^*}{e^*}\vec{v_s} \right]\cdot d\vec{l} + \phi
 = n \phi_0
 \label{fluxoid}
\end{equation}

\noindent where $m^*$, $e^*$ and $\vec{v_s}$ are respectively the mass, the charge and the velocity of the supercurrent carriers, $\phi$ is the magnetic flux threading the integration contour and $n$ is an integer. The integer number $n$ of fluxoid quanta in the ring is a good quantum number and can be seen as the number of vortices threading the ring, each vortex carrying a fluxoid quantum $\phi_0$. If this kind of systems has been widely studied theoretically \cite{vodo03,baelus01} and experimentally \cite{mosh} from an electrical point of view \cite{zhang97,morelle}, by magnetic decoration \cite{pannet} or through magnetization measurements \cite{pedersen,geim}, little is known about the thermal behaviour of superconducting mesoscopic loops\cite{bourgprl,ongPRB2006,gandit,buisson,deo}. Here we report highly sensitive heat capacity measurements performed on an array of independent mesoscopic superconducting rings of size comparable to the superconducting coherence length $\xi(T)$ under an applied magnetic field. 

In a previous paper (see Ref.\cite{bourgprl}), we have already demonstrated that multiple phase transitions between states with different vorticities are accompagnied by discontinuities of the heat capacity. Each vortex entrance (or expulsion) is associated to a mesoscopic phase transition from the $n$ to the $n+1$ (or $n-1$) state, and as the magnetic field is increased, an oscillating heat capacity is measured with a periodicity corresponding to a flux quantum $\phi_0$ threading a loop. Under specific conditions (lowest temperatures, zero field cooling) it was also shown that several vortices could enter or exit the loop at the same time, leading to oscillations of the heat capacity with a periodicity of 2$\phi_0$ at 0.85 K and 3$\phi_0$ at 0.70 K. An important feature distinguishing this article from Ref. \cite{bourgprl} is that here we focus only on the 1$\phi_0$ periodic oscillations of the heat capacity: even at lowest temperatures the system is prepared to have only a 1$\phi_0$ periodic component. Indeed as it was already mentionned in Ref.\cite{bourgprl}, the magnetic history of the loops is very important and determines the way the sample behaves. If the magnetic field is swept from -40 mT down to zero even at 0.6 K the modulation of the heat capacity is $\phi_0$ periodic. The 3$\phi_0$ periodic signal as shown in Ref.\cite{bourgprl} is evidenced \emph{only} when the loops are cooled in zero field before the magnetic field is swept from zero to a high value of H. This hysteresis of the heat capacity of mesoscopic sample has been observed on many different situations; this behaviour will be discussed in a forthcoming paper. Here we avoid the appearance of 2$\phi_0$ and 3$\phi_0$ periodic oscillations, in order to study the fine evolution of the periodicity of the heat capacity jumps with regard to the varying magnetic field and to its sweeping direction. The advantages of working at low temperature are the enhancement of the signal-to-noise ratio and the increase of the total number of vortices a loop can host. We stress here that in Ref.\cite{bourgprl} we studied the temperature and magnetic field ranges enabling the appearance of multiquanta transitions. It was shown that the pseudoperiod of the heat capacity versus magnetic flux $C(H)$ can take \emph{discrete} values $n\times \phi_0$. In the present paper we focus on the $1\phi_0$ regime and show that the pseudoperiod of $C(H)$ is \emph{continuously} changing as $H$ is swept.

\section{Experimental results and analysis}
\subsection{Heat capacity measurements}
 
The sample studied in this work is composed of an array of $N=4.5 \times 10^{5}$ identical non interacting superconducting aluminum square loops (see the inset of Fig.\ \ref{fig2}: each loop has 2 $\mu$m side,
$w = 230$ nm arm width, $d = 40$ nm thickness, for a total
mass $m = 80$ ng of aluminum). The separation of neighboring loops is 2 $\mu$m.
The inter-loop interaction is neglected; we indeed do not expect it to affect our results significantly because 
for our sample the mutual inductance ($\simeq -20$ fH) is much smaller than the
self inductance ($\simeq 5$ pH). Thus the mutual magnetic flux always remains much smaller than $\phi_0$,
and the energy of magnetic interaction between loops is much smaller than the free energy
of a single loop. We will consider in the following that all the thermal signals are additive, hence the measured heat capacity is $N$ times the heat capacity of a single loop.

The $N$ mesoscopic square loops are patterned by electron beam lithography on a home-made specific heat sensor and aluminum is deposited by thermal evaporation. The sensor \cite{fom97,bourgprl} is composed of a large (4 mm$\times$4 mm) and very thin (5 $\mu$m) silicon membrane suspended by twelve silicon arms. On this membrane, a copper heater and a NbN thermometer are deposited through regular photolithography. Resistances of these thin film transducers are measured by a four point probe technique. In the case of the Cu heater we can thus measure the power injected in the calorimeter. In the case of the thermometer the four wire measurement reduces the measuring noise, enabling the read out of very small temperature variations on which our specific heat measurement is based. Indeed using this device the total heat capacity can be measured by ac-calorimetry \cite{sullivan}. The principle of this technique lies in applying a sinusoidal current in the heater at the frequency $f$, resulting in oscillations of the temperature of the membrane at twice this frequency. Thus, when measuring the voltage response on the dc-biased thermometer one gets the amplitude $\delta T_{\rm ac}$ of the oscillations of the temperature and hence information about the thermal properties (thermal conductivity and heat capacity) of the silicon membrane and the nanosystems it contains. This method has been largely described in numerous publications \cite{bourgprl,sullivan,riou97,fom97}. The calorimetry setup is cooled down to 0.55 K using a $^3$He cryostat. 

The correct working frequency is obtained by measuring the response function $f \times \delta T_{\rm ac}$ of the sensor. When this response is extremal, i.e. becomes independent of the frequency $f$, the system can be considered as quasiadiabatic and then $\delta T_{\rm ac}$ is only related to the heat capacity and to known parameters (frequency $f$ and injected power) \cite{sullivan}. We define the \emph{adiabatic plateau} as the frequencies interval such as $f \times \delta T_{\rm ac}$ is greater than $99\%$ of its maximum. In our case we obtain $f \in [108;149 \rm ~Hz]$ at 0.6 K; in these conditions the thermal excitation is faster than the characteristic time of the heat loss to the thermal bath but slower than the heat diffusion time in the system (sensor and sample). The major advantage of working with lithographied loops is the high thermal contact between the nano-objects and the silicon sensor. In this quasiadiabatic limit, all superconducting loops can be considered as being at the same temperature. The amplitude $\delta T_{\rm ac}$ of the oscillations of temperature can be tuned between 1 mK to 15 mK depending on the working temperature and on the resolution needed for the measurements. The ac-calorimetry enables averaging of the measured signal, thus reducing the error bar for each measured data point. Typically by averaging over 10 seconds, this apparatus allows measurements of heat capacity within 10 femto-Joule per Kelvin, which corresponds to an energy sensitivity as small as few atto-Joule (10$^{-18}$ Joule).

\begin{figure}
\centering
\includegraphics[width=12cm,angle=0]{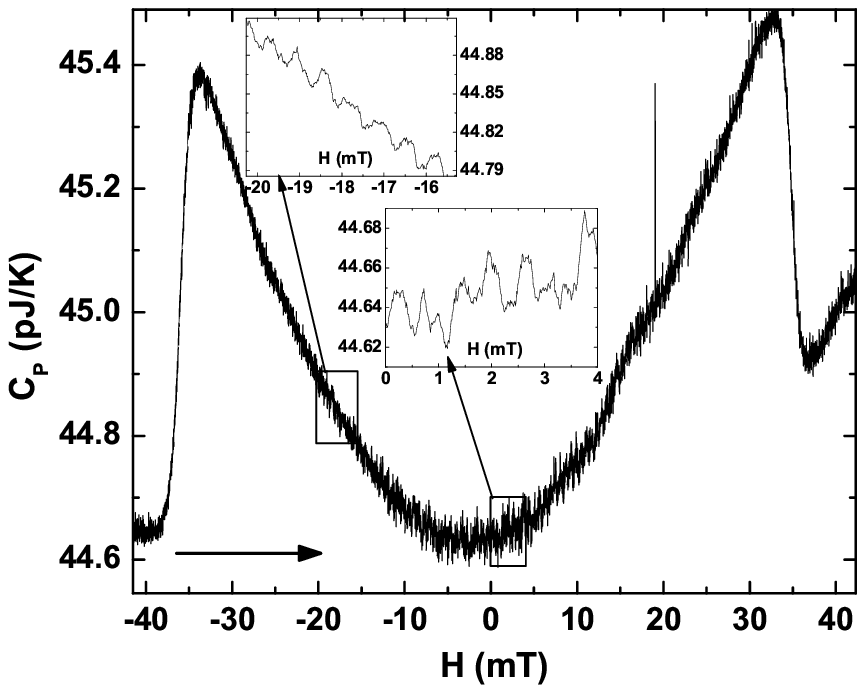}
\caption{
Heat capacity measurement of the superconducting rings at 0.75 K. The heat capacity jumps at -35 mT and 35 mT corresponds to the transition between the superconducting state and the normal state. The insets focus on two areas after signal processing reducing the visible noise (10 points adjacent averaging): small heat capacity oscillations appears to be superimposed on a slowly varying background. The arrow indicates that in this experiment the field is swept from -45 mT to 45 mT.}
\label{fig1}
\end{figure}

\begin{figure}
\centering
\includegraphics[width=12cm,angle=0]{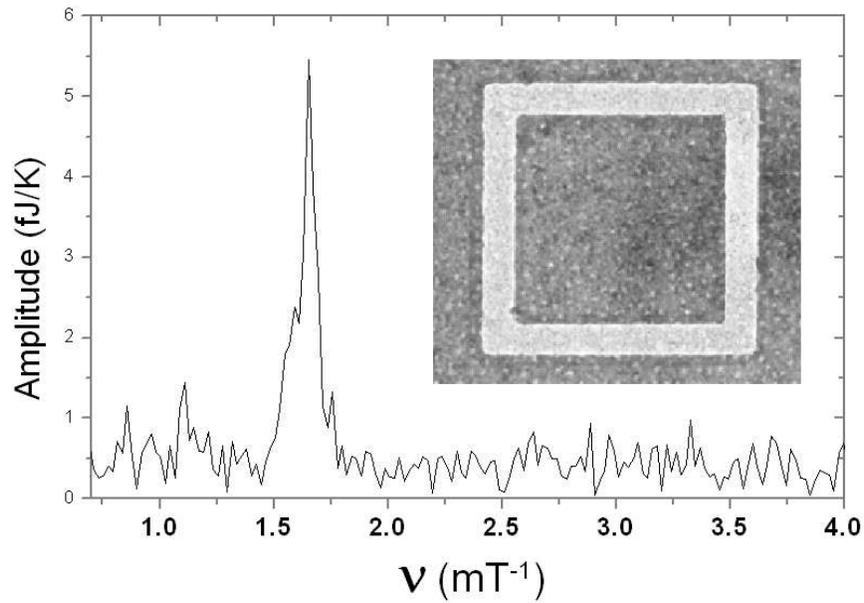}
\caption{
Modulus of the Fourier transform of the signal shown in Fig. 1 windowed from H=-30 mT to H=30 mT (superconducting part of the plot), after low pass filtering to remove the background. A large peak appears at the frequency $\nu=$ 1.65 mT$^{-1}$, which corresponds to one magnetic flux quantum in a square of 1.86 $\mu$m side. The inset shows an electron micrograph of one loop : the external side is 2.05 $\mu$m and the internal side is 1.65 $\mu$m.}
\label{fig2}
\end{figure}

The sensor is placed in the center of a large superconducting coil so that a tunable magnetic field $\vec{H}$ can be applied perpendicular to the plane of the loops. The area containing the loops is 2.5 mm$\times$2.8 mm large, and the difference between the field on the axis and the field at the border of the sample has been calculated to be less than 0.1$\%$. This is regarded as negligible and hence the field is considered as homogenous all over the sample surface. A heat capacity measurement at 0.75 K is shown on Fig. \ref{fig1}. The measurement is performed under a perpendicular magnetic field $\vec{H}$ swept from -45 mT to 45 mT at constant temperature. Two major signatures of second order phase transition from the normal phase to the superconducting phase can be seen at -35 mT and 35 mT through a heat capacity jump of 0.75 pJ/K. Apart from these significant changes in the heat capacity, subsignatures appear at intermediate magnetic fields. As it is expanded in the insets of the Fig. \ref{fig1}, oscillations of the heat capacity can be observed with a period of approximately 0.6 mT. This magnetic field interval corresponds to one superconducting flux quantum $\phi_0$ ($\phi_0=\frac{h}{2e})$ threading a square of 1.86 $\mu$m side, a contour which is included in the volume of a single aluminum loop (see the inset of Fig. \ref{fig2}). As it was demonstrated earlier\cite{bourgprl}, each vortex entrance (or expulsion) changes the vorticity $n$ by one, leading to a phase transition characterized by a heat capacity jump. As the flux through a loop is swept such phase transitions occur with a periodicity close to $\phi_0$ and hence an oscillatory behavior appears. The amplitude of these oscillations ($\approx$ 20 fJ/K) is of the order of the measuring noise ($\approx$ 10 fJ/K by averaging over 10 seconds). So the oscillations are hidden in the noise when looking directly at the raw data of Fig. \ref{fig1}. Oscillations appear more clearly (as shown in the insets of Fig. \ref{fig1}) by smoothing the raw signal through a 10-points adjacent-averaging process. 

In order to better exhibit the oscillatory behavior, we perform a filtering of the background of the signal (general trend corresponding to low frequencies); then we calculate a Fourier transform of the oscillating signal isolated in this way. The result is presented in Fig. \ref{fig2} where the modulus of the Fourier transform shows the oscillating contribution of the superconducting loops inside the superconducting area (H = -30 mT to H = 30 mT). A large peak is exhibited at the frequency $\nu=1.65$ mT$^{-1}$, but not much information can be extracted from this broad peak. Indeed since our aim is to study how vortices entrances or expulsions affect the period of the signal as the field is swept, this single spectral representation is not adapted. In fact, to extract relevant information of a non stationary signal, one needs to have a signal processing tool with a high resolution in both magnetic field $H$ and frequency domain ($\nu$). Hence we will use a specific time-frequency representation, the Wigner-Ville distribution.

\subsection{Signal processing}

The magnetic field interval between two successive heat capacity jumps seems to evolve as the field is increased. This new feature, not mentioned in Ref. \cite{bourgprl}, is quite interesting if one wants to deeper understand the specific character of the metastability of the thermodynamic state in a superconducting ring. From electrical measurements on a mesoscopic superconducting disk, Baelus {\it et al.} have shown that the penetration and expulsion fields can depend on the measurement temperature \cite{baelus05}. Here we show that the penetration and expulsion fields depend on the magnetic history of the sample as well, i.e. they depend on the applied magnetic field and on its sweeping direction. In order to study how the periodicity of the $C_p(H)$ curves evolves as the magnetic field is varied, we use the Wigner-Ville distribution (WVD)\cite{signproces}, a time-frequency processing tool adapted to study non-stationary signals and giving a better resolution in frequency than a regular Fourier transform, especially when the characteristic frequency of the signal is changing continuously with time. For our data the role of time is played by the magnetic field.

The main purpose of the WVD is to image the energy distribution of a non-stationary signal in the time-frequency space. Because any tool based on simple Fourier transform needs to find an appropriate balance between time and frequency resolution, the WVD solves this problem by calculating

\begin{equation}
W_{C_p}(H,\nu)=\int^{+\infty}_{-\infty} C_p (H+\frac{\tau}{2})C_p(H-\frac{\tau}{2})e^{-2i\pi\nu\tau}d\tau
\label{WVD}
\end{equation}

\noindent where $C_p(H)$ is the heat capacity measured versus magnetic field.
The WVD gives the best compromise between temporal and frequency resolutions. However the WVD processing method has a disadvantage: the distribution is perturbed by the presence of cross-terms. These cross-terms emerge from interferences between characteristic frequencies. To resolve that problem, instead of using the raw definition given in Eq. (\ref{WVD}) we use the smoothed pseudo Wigner-Ville distribution (SPWVD) given by

\begin{eqnarray}
\begin{array}{cc}
S_{C_p}(H,\nu)= \int^{+\infty}_{-\infty} \left| h(\frac{\tau}{2}) \right|^2 \times \\
\left[ \int^{+\infty}_{-\infty} 
g(s-H) C_p(s+\frac{\tau}{2})C_p^{*}(s-\frac{\tau}{2}) ds \right] 
e^{-2i\pi\nu\tau} d\tau 
\label{PWVD}
\end{array}
\end{eqnarray}

where $h(\tau)$ is a short time window allowing  a spectral smoothing to reduce cross-terms between shifted time terms and $g(s-H)$ is a frequency window allowing a time smoothing to reduce cross-terms between shifted frequency terms. Several kinds of windows can be used (see Ref.\ \cite{harris}). In the same way as in classical spectral analysis, boxcar window function are prohibited, so we use in this work smoothing in time and frequency through Hanning windows \cite{harris}. By applying two Hanning windows ($h$ and $g$) one gets the same result as a 2D convolution of the WVD by a 2D window. Consequently SPWVD causes a slight loss of resolution for the characteristic frequency but enables a strong attenuation of the cross terms.

\begin{figure}
\centering
\includegraphics[width=12cm,angle=0]{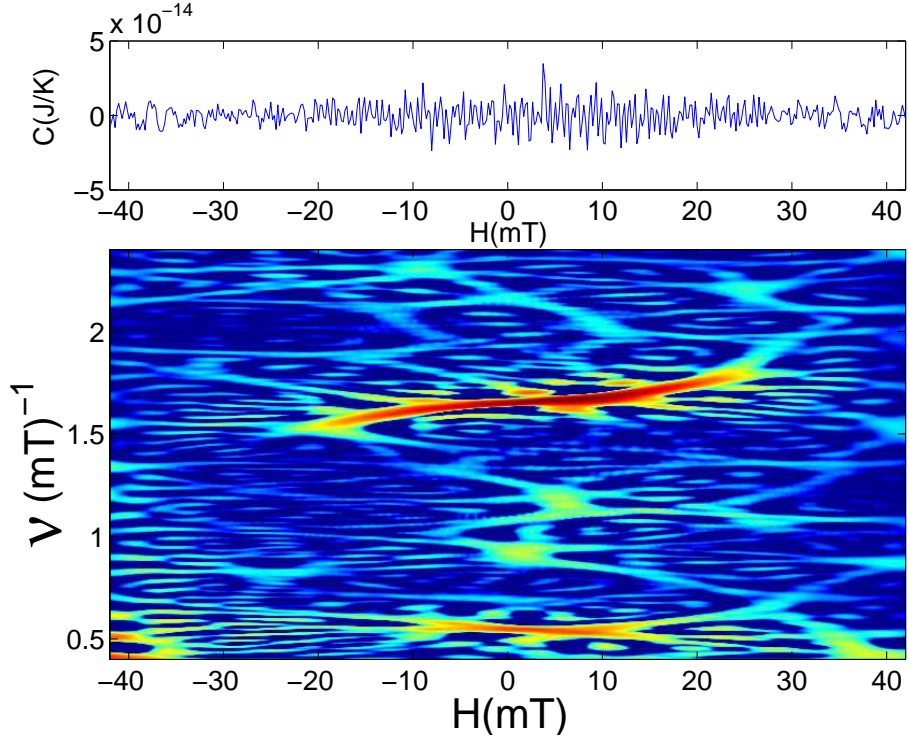}
\caption{
(Color online)
This smoothed pseudo Wigner-Ville distribution (SPWVD) is calculated from the signal presented in Fig. 1 after low pass filtering to remove the slowly varying background. The upper part of the graph shows the heat capacity signal after substraction of that trend. The lower part is the SPWVD of the signal (a magnetic field-frequency representation of the heat capacity data). The evolution of the periodicity versus the magnetic field appears clearly in the middle of the graph, where the frequency $\nu$ of the oscillations varies from 1.50 mT$^{-1}$ to 1.78 mT$^{-1}$, which represents more than a 15\% change. The color code of this graph is only used to illustrate the difference in the calculated amplitude: from blue for the lowest amplitudes to red for the highest ones.}
\label{fig3}
\end{figure}

We use the SPWVD to analyse our $C_p(H)$ signal. An example is shown in Fig. \ref{fig3}. This is a density plot of $|S|$ in the $H-\nu$ plane; this SPWVD is calculated from the $C_p(H)$ data presented on Fig. \ref{fig1}. At a given magnetic field $H$ (horizontal axis) the amount of power contained in the harmonic with frequency $\nu$ (read on the vertical axis) is given by a colour code, from the deep blue for the lowest amplitudes to the red for the highest ones. No quantitative physics will be extracted from the amplitudes given by the color code, which is only used to localize the characteristic frequency read on the vertical axis. Roughly speaking, for a given field $H$, the ordinate of the red area is the local frequency of the signal in the vicinity of $H$, i.e. the inverse of the local periodicity of $C_p(H)$. Thus we clearly get the evolution of the periodicity in the vortex expulsion area (negative magnetic field) and in the vortex penetration area (positive magnetic field): at -25 mT the local frequency is (1.50$\pm$0.04) mT$^{-1}$ (i.e. the period is 0.67 mT) and at 25 mT the frequency reaches (1.78$\pm$0.02) mT$^{-1}$ (period = 0.56 mT). The change of period of the heat capacity oscillations is thus above 15\%. This is an illustration that all the jumps in the heat capacity observed in that kind of experiment are not directly related to one $\phi_0$, but are linked to the metastability of the thermodynamic states which will change with the applied magnetic field, and so with the number $n$ of vortices in the loops. Either the periodicity of the jumps is not exactly an integer number of quantum magnetic flux $\phi_0$, or the contour through which the fluxoid is strictly quantized evolves with the magnetic field. This will be further discussed in the theory part of the article.

\begin{figure}
\centering
\includegraphics[width=12cm,angle=0]{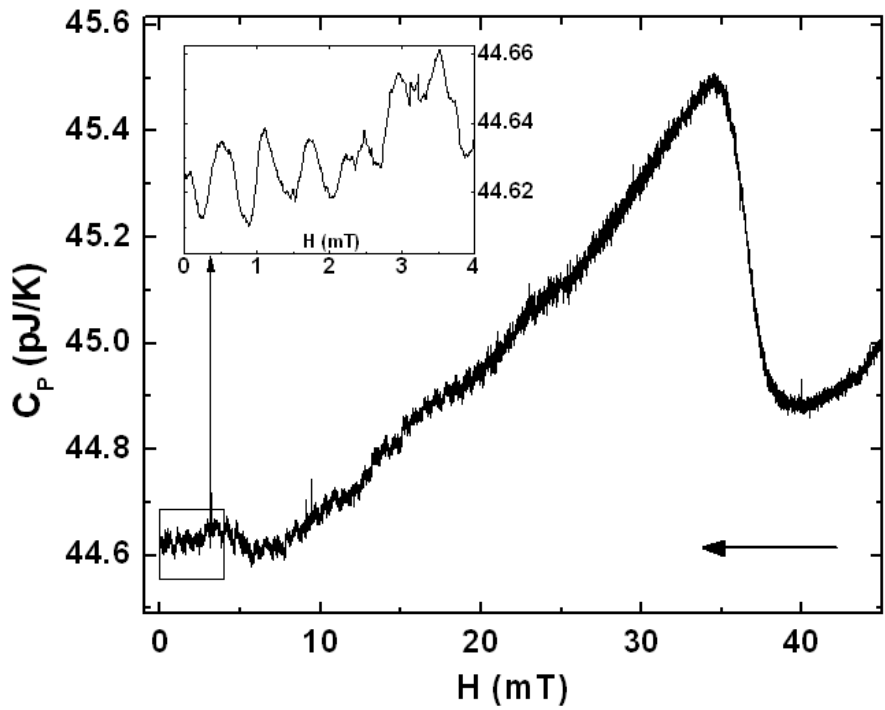}
\caption{
Heat capacity measurement of superconducting rings at 0.75 K in decreasing magnetic field. The oscillations with a periodicity of one $\phi_0$ are expanded in the inset. The arrow indicates that in this experiment the field is swept from +45 mT to 0 mT.}
\label{fig4}
\end{figure}

The non-stationarity of the signal does not depend on the temperature; closer to $T_c$ the same observation can be made. It does not depend neither on the geometry because the same results have been obtained on circular superconducting rings of 1 $\mu$m and 2 $\mu$m in diameter. The crucial parameter is the sweeping direction of the applied magnetic field. If the absolute value $|H|$ of the field is decreased, the sample is in the situation where vortices are expelled outside the loops. The opposite behavior is obtained when vortices are penetrating the loops when $|H|$ is decreased. We clearly illustrate that point in Fig. \ref{fig4} and Fig. \ref{fig5}, (the analogues of Fig. \ref{fig1} and Fig. \ref{fig3}) where the magnetic field is decreased from 40 mT. In this case, as the field is decreased, the frequency of the $C_p$ jumps is increasing. This evidences the fact that the sweeping orientation of the field determines the evolution of that periodicity. The magnetic field sweep rate was the same ($\approx$4.5 mT per hour) when getting the data points of Figs. \ref{fig1} and \ref{fig4}, and it can be varied between 1 mT/hour to 20 mT/hour without affecting the results.

\section{Theory}

A reasonably good description of our experimental observations can be obtained in the framework of the Ginzburg-Landau (GL) theory of superconductivity \cite{tink04}. In this theory the free-energy density in the superconducting state $f$ is written as:
\begin{eqnarray}
f(\vec{r}) &=& f_{\mathrm{normal}}(\vec{r}) + \alpha(T) \left| \psi(\vec{r}) \right|^2 + \frac{\beta(T)}{2} \left| \psi(\vec{r}) \right|^4
\nonumber \\ 
&+& \frac{1}{2 m^{*}} \left| \left[ \frac{\hbar}{i} \vec{\nabla} - \frac{e^{*}}{c} \vec{A} \right] \psi(\vec{r}) \right|^2
\label{fs}
\end{eqnarray}
where $f_{\mathrm{normal}}$ is the free-energy density in the normal state that is assumed to be independent of the magnetic
field $\vec{H} = \mathrm{rot} \vec{A}$, $\psi(\vec{r})$ is the complex order parameter,
$m^{*}$ and $e^{*}$ are the effective mass and charge of the carriers \cite{tink04}.
We assume $m^{*} = 2 m$ and $e^{*} = 2 e$, where $m$ and $e$ are the electron mass and charge, respectively, because the
charge carriers in a superconductor are Cooper pairs.
The coefficient $\alpha(T)$ changes sign at the superconducting transition $T = T_c$, while
the coefficient $\beta(T)$ is always positive and depends on temperature only weakly (if at all).
Equation (\ref{fs}) should be integrated over the volume of the superconducting sample (loop in our case) to obtain its free
energy $F$. For simplicity, we will consider a circular ring with a diameter $R$ and the same average perimeter as the square loops studied in our
experiments. This will allow us to obtain at least a qualitative understanding of the experimental results without too complicated mathematics \cite{fomin98}. When the magnetic field $\vec{H}$ is perpendicular to the plane of the ring, i.e. directed
along the $z$ axis, $\vec{A} = \vec{e}_{\varphi} H \rho/2$.
In addition, if the thickness $d$ of the ring is much smaller than the magnetic field penetration depth $\lambda(T)$, the magnetic field $\vec{H}$ can be considered equal to the external
(applied) field.

Minimizing the free energy $F$ by variational methods leads to the famous
GL differential equation for $\psi(\vec{r})$ \cite{tink04}. In Ref.\ \cite{bourgprl} we have solved this equation numerically and
have shown that this provides at least qualitative description of the experimental measurements. Here we will show that
the problem can be treated analytically as well, provided that certain additional but reasonable assumptions are
adopted. The central assumption is that both $d$ and the width of the ring arm $w$ are smaller than the GL
coherence length $\xi(T)$. This allows us to neglect $z$- and $\rho$-dependences of the order parameter $\psi(\vec{r})$ and write $\psi(\vec{r}) =  \psi(\varphi)$. The free energy $F$ is then minimized by
$\psi_n(\vec{r}) = \left| \psi_n \right| \exp(i n \varphi)$ with integer $n$ and \cite{zhang97}

\begin{eqnarray}
\left| \psi_n \right|^2 &=& -\frac{\alpha}{\beta} - \frac{1}{\beta} \frac{\hbar^2}{4 m R^2}
\left[ \left(n - \phi \right)^2 +
\left( \frac{n^2}{3} + \phi^2 \right) \left( \frac{w}{2 R} \right)^2
\right]
\label{psin}
\end{eqnarray}

Here we assumed that $w$ is much smaller than the average radius $R$ of the ring and kept only the leading terms in $w/R$;
$\phi$ stands for $\Phi/\Phi_0$ with $\Phi = H \pi R^2$ the magnetic flux through the ring and
$\Phi_0 = h/2 e$ the magnetic flux quantum.
The state $\psi_n(\vec{r})$ is called `$n$-giant vortex state' because in this state, the ring hosts
$n$ magnetic vortices sharing the same core; we will also refer to it as `$n$-vortex state' or simply `state $n$' for brevity. 
The free energy $F_n$ of the ring in the $n$-vortex state is:

\begin{eqnarray}
F_n &=& -V \frac{\alpha^2}{2 \beta} 
\left\{ 1 - \left[\frac{\xi(T)}{R} \right]^2 \right.
\times \left. \left[\left(n - \phi \right)^2 + \left( \frac{n^2}{3} + \phi^2 \right) \left( \frac{w}{2 R} \right)^2
\right] \right\}^2
\label{freen}
\end{eqnarray}\noindent where $V = 2 \pi R w d$ is the volume of the ring and we introduced the GL coherence length
$\xi(T)^2 = \hbar^2/4 m \left| \alpha(T) \right|$.

By noting that $\alpha^2/2 \beta = \mu_0 H_c(T)^2/2$ with $H_c(T)$ the thermodynamical critical magnetic field \cite{tink04}
and by using the empirical expressions \cite{tink04}
$H_c(T) = H_c(0) (1 - t^2)$,
$\xi(T)^2 = \xi(0)^2 (1 + t^2)/(1 - t^2)$ with $t = T/T_c < 1$,
we can put Eq.\ (\ref{freen}) in the following form:

\begin{eqnarray}
F_n &=& -V \frac{\mu_0 H_c(0)^2}{2} (1 - t^2)^2
\left( 1 - \gamma_n \frac{1 + t^2}{1 - t^2} \right)
\label{freen2}
\end{eqnarray}

\noindent where $\gamma_n = [\xi(0)/R]^2 [(n - \phi)^2 + ({n^2}/3 + \phi^2)(w/2R)^2]$.
From the condition $F_n = 0$ with $\phi \simeq n$ we estimate the maximum number $n_{\mathrm{max}}$ of magnetic vortices that can be hosted by a ring to be
$n_{\mathrm{max}} \simeq \sqrt{3} R^2/w \xi(T)$.
For the rings with parameters corresponding to our experiments ($T_c=$1.18 K, $\xi(0)=$ 130 nm, $w=$230 nm, $R=$ 1.13 $\mu$m), we find
$n_{\mathrm{max}} =$ 27, 37, and 43 at $T=$ 1.00 K, 0.85 K, and 0.75 K, respectively.
 
The heat capacity $C_p^{(n)}$ of the ring in the $n$-vortex state is obtained by differentiating $F_n$
twice with respect to the temperature:

\begin{eqnarray}
C_p^{(n)} &=& -T \frac{\partial^2 F_n}{\partial T^2}
= V \frac{2 T \mu_0 H_c(0)^2 }{T_c^2}
\left[ \gamma_n^2 - 1 + 3 t^2 (1 + \gamma_n)^2 \right]
\label{cn}
\end{eqnarray}

We plot the heat capacity of the rings at $t=$ 0.95 (such a choice will be explained in the following) in different $n$-vortex states as a function of normalized magnetic flux $\phi$ in Fig.\ \ref{theory_fig1} by dashed lines [lines are the same in both panels].
When the magnetic field is varied continuously while the temperature is kept fixed,
(as in our experiments), the ring can exhibit transitions between states with different $n$. In an increasing magnetic field, for example, the transition from $n$- to $(n+1)$-vortex state becomes energetically favorable when
$F_{n}$ becomes larger than $F_{n+1}$. The points $\phi_n$ where these (phase) transitions occur can be therefore
found from the condition $F_n = F_{n+1}$. This yields
$\phi_n = (n + 1/2)(1 + w^2/12 R^2)$, which appears to be exactly the points where $C_p^{(n)}$ and $C_p^{(n+1)}$ specific
heat curves cross.
In the thermodynamic equilibrium, therefore, the heat capacity $C_p(\phi)$ should follow the lowest of the dashed curves in
Fig.\ \ref{theory_fig1}, thus exhibiting periodic variations with magnetic field with a period
$\phi_{n+1} - \phi_n = 1 + w^2/12 R^2 \approx 1$, similar to what we observe in the experiments.

A closer inspection of experimental results (Figs.\ \ref{fig1} and \ref{fig4} and Ref.\ \cite{bourgprl}) reveals that in some of our measurements the peaks of $C_p(H)$ appear more asymmetric than expected from the theory in the thermodynamic equilibrium, whereas in other measurements this asymmetry is not that striking. This suggests that at least in a part of our experiments the thermodynamic equilibrium is not reached. The rings exhibit a sort of non-equilibrium phase transitions, remaining in a state with a given number $n$ of magnetic vortices beyond the `critical' magnetic
field $\phi_n/\pi R^2$, where the thermodynamic transition to the $(n+1)$-vortex state should occur. To study such a
possibility, we analyze the stability of the $n$-vortex state $\psi_n(\vec{r}) = \left| \psi_n \right| \exp(i n \varphi)$ 
with respect to an admixture of a $k$-vortex state $\psi_k(\vec{r}) = \left| \psi_k \right| \exp(i k \varphi)$.
Following Bezryadin \textit{et al.} \cite{buzdin95}, we write the free energy corresponding to the state
$\psi(\vec{r}) = \psi_n(\vec{r}) + \eta \psi_k(\vec{r})$ as

\begin{eqnarray}
F(\eta) = F_n + A \eta^2 + B \eta^4
\label{fstab}
\end{eqnarray}

\noindent where $A = V \left| \psi_k \right|^2
[ \alpha + 2 \beta \left| \psi_n \right|^2 + \gamma_k \left| \alpha \right| \xi(T)^2/\xi(0)^2 ]$
and $B = \beta \left| \psi_k \right|^4/2$.
In contrast to Ref.\ \cite{buzdin95}, we do not take the limit $w  \rightarrow 0$ in the above
expressions and keep the terms of order $(w/R)^2$ (but not the higher-order ones because our analysis is
still limited to $w \ll R$).

Because $B > 0$, for $A > 0$ the minimum of $F(\eta)$ is reached at $\eta = 0$. This corresponds to the situation when the $n$-vortex state is stable and any admixture of a
different, $k$-vortex state is energetically unfavorable. In contrats, if $A < 0$, the minimum of
Eq.\ (\ref{fstab}) is reached for $\eta = \pm (-A/2B)^{1/2} \ne 0$. In this case the admixture of the
$k$-vortex state is
energetically favorable and the $n$-vortex state becomes unstable. The condition $A = 0$ defines therefore an instability threshold (or a `superheating boundary' \cite{buzdin95,horane96}) of the state $n$.

Let us now apply the above stability analysis to our experimental situation in which the magnetic field is slowly varied and transitions between different giant vortex states are observed.
It is reasonable to assume that the transition from a state $n$ to a state $k$ takes place when the former
becomes unstable with respect to an admixture of the latter. Since in our experiments only transitions
from $n$ to $n + 1$ (or to $n - 1$, in decreasing field) states are observed, we put $k = n \pm 1$
and find that the $n$-vortex state becomes unstable if $\phi$ is increased above (decreased below)

\begin{eqnarray}
\phi_n^{\pm} &=&
n \mp 1 \pm \sqrt{2 + [R/\xi(T)]^2} 
\nonumber \\
&\pm&
\left(\frac{w}{2 R} \right)^2
\times
\left\{1  - \frac{4/3 + [R/\xi(T)]^2/2}{\sqrt{2 + [R/\xi(T)]^2}}
\right.
\mp \left. n \left[ 1 \pm \frac{2 n /3 \mp 4/3}{\sqrt{2 + [R/\xi(T)]^2}} \right]
\right\}
\label{phipm}
\end{eqnarray}

The instability of the $n$-vortex state in an increasing magnetic field occurs at larger fields than the thermodynamic transition does, i.e.
$\phi_n^{+} > \phi_n$. Similarly, in a decreasing field one has: $\phi_n^{-} < \phi_n$ (note that Eq.\ (\ref{phipm}) reduces to the result of Ref.\ \cite{buzdin95} if we set $w = 0$). This suggests that in an experiment the ring can remain in the state $n$ (which in this case becomes metastable) beyond the thermodynamic transition and up to the instability point where a jump to $n \pm 1$ state occurs. This is illustrated by solid lines in Fig.\ \ref{theory_fig1}. These lines are obtained by following a given $C_p^{(n)}$ curve up to $\phi_n^{\pm}$ given by Eq.\ (\ref{phipm}) and then jumping to the next $C_p^{(n \pm 1)}$ curve. The resulting dependence $C_p(\phi)$ is close to the experimental one (Figs.\ \ref{fig1} and \ref{fig4}). 
On Fig. \ref{theory_fig1} the temperature was set to $t=0.95$. A lower temperature would have lead to multiquanta transitions (see Ref. \cite{bourgprl}). In our experiments - given a specific preparation of the sample as explained in the first section of this paper - we force the system to remain in the $1\phi_0$ oscillatory regime even at low temperatures. Thus to compare our signals with the model we used in Fig. \ref{theory_fig1} a temperature leading to a $1\phi_0$ regime.


Equation (\ref{phipm}) can also be used to study the periodicity of the $C_p(H)$ dependence in more detail. The distance between consecutive instability points [jumps of $C_p(\phi)$] in the increasing magnetic field,
$\Delta \phi_n^{+} = \phi_n^{+} - \phi_{n-1}^{+}$, and in the decreasing field,
$\Delta \phi_n^{-} = \phi_{n+1}^{-} - \phi_{n}^{-}$, are found to be:

\begin{eqnarray}
\Delta \phi_n^{\pm} &=&
1 - \left( \frac{w}{2 R} \right)^2
\left[1  - \frac{2 \mp 4 n/3}{\sqrt{2 + [R/\xi(T)]^2}} \right]
\label{dphipm}
\end{eqnarray} 

\begin{figure}
\centering
\includegraphics[width=12cm,angle=0]{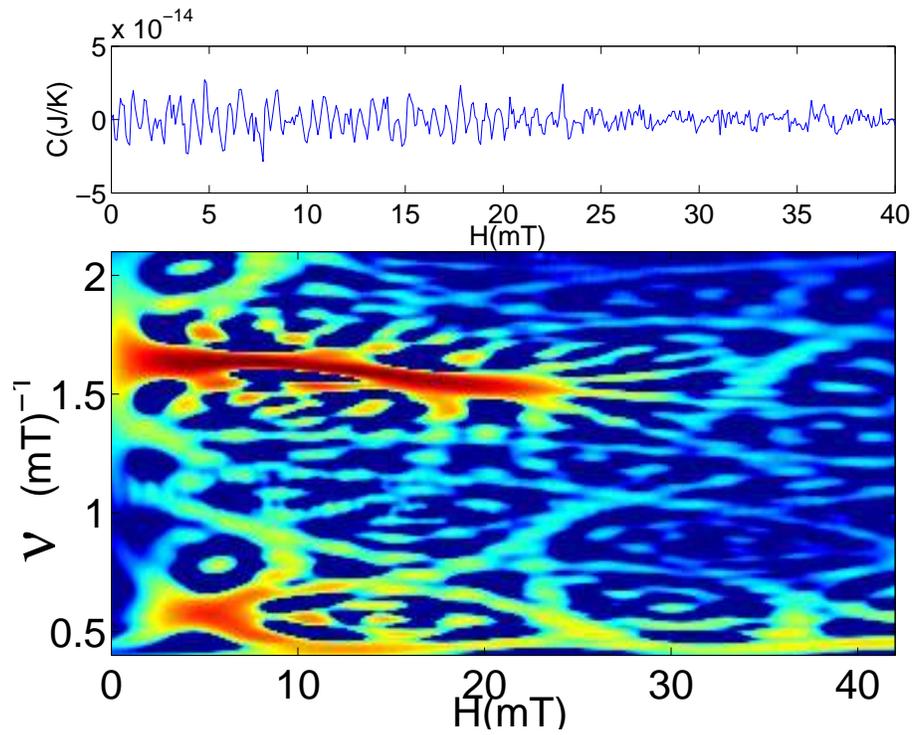}
\caption{
(Color online)
This SPWVD is calculated from the signal presented in Fig. \ref{fig4} after low pass filtering to remove the slowly varying background (upper part of the graph). The lower part is the SPWVD of that signal. The evolution of frequency versus magnetic field is inverted as compared to the case where the magnetic field is increased (see Fig. \ref{fig3}). The color code is the same as in Fig. \ref{fig3}.}
\label{fig5}
\end{figure}

\begin{figure}
\centering
\includegraphics[width=12cm,angle=0]{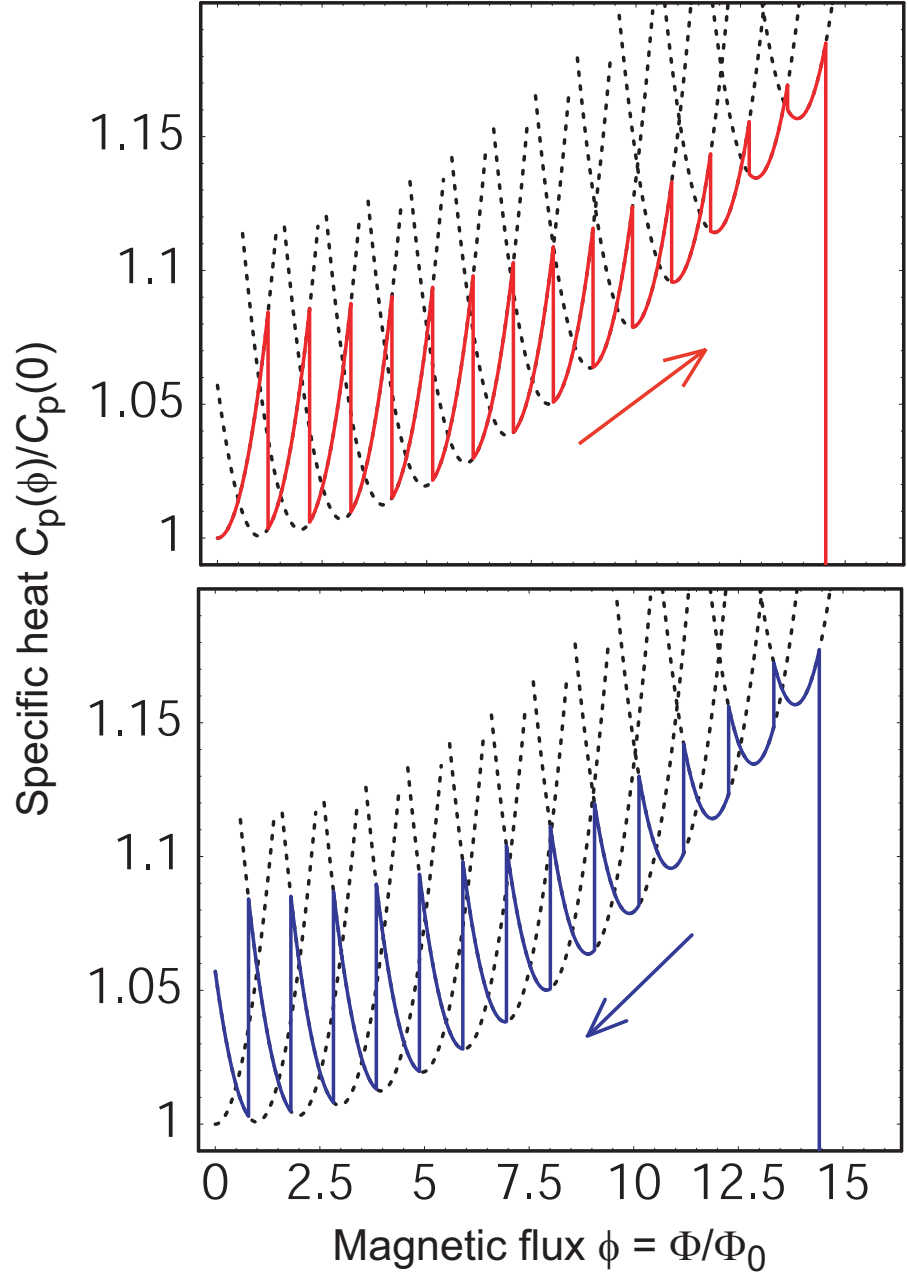}
\caption{
(Color online)
Heat capacity of a superconducting ring calculated from the Ginzburg-Landau theory at $T/T_c = 0.95$.
The ring is assumed to be circular with the same average perimeter as the square loop studied in the
experiments and with the same arm width and thickness.
Dashed lines correspond to giant vortex states with vorticity $n$.
Transitions between consecutive $n$-vortex states taking place at `critical' magnetic fluxes given by
Eq.\ (\ref{phipm}), yield the (red) solid curve in the upper panel in the increasing magnetic field
and the (blue) solid curve in the lower panel in the decreasing magnetic field.
The large jumps of the heat capacity at $\phi \simeq 14.5$ correspond to the normal-superconducting
transition.}
\label{theory_fig1}
\end{figure}

\begin{figure}
\centering
\includegraphics[width=12cm,angle=0]{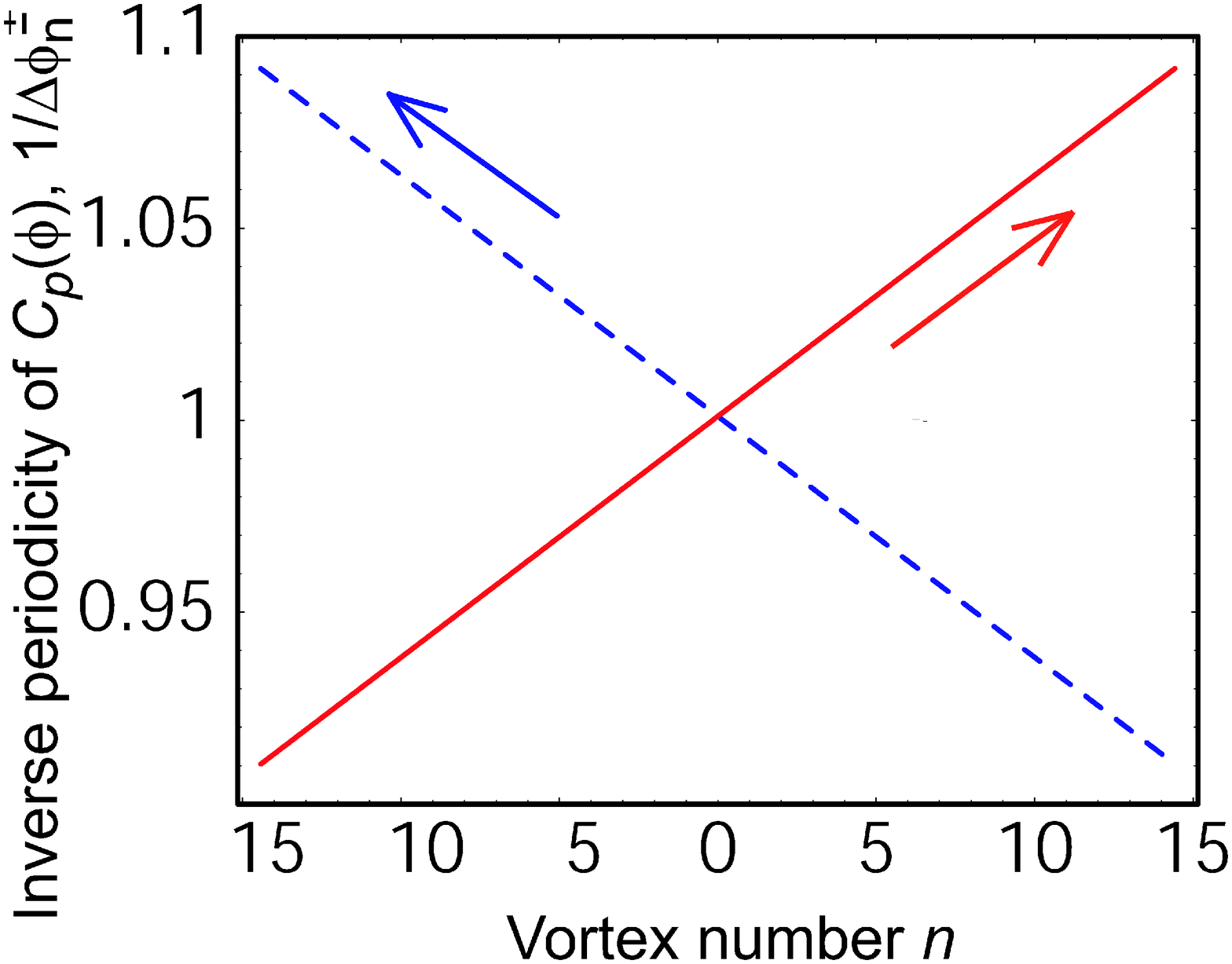}
\caption{
(Color online)
Inverse periodicity of the heat capacity $C_p(\phi)$ [Eq.\ (\ref{dphipm})] at $T/T_c = 0.95$
as a function of the magnetic vortex
number $n$. The inverse periodicity increases with $n$ in the increasing magnetic field (red solid line) and decreases
with $n$ in the decreasing magnetic field (blue dashed line). The tendency is the same as found experimentally
(Figs.\ \ref{fig3} and \ref{fig5}, respectively). The magnitude of the effect $\sim 10$\% is also comparable to the
measured one.}
\label{theory_fig2}
\end{figure}

\noindent We see that the periodicity of heat capacity jumps is not constant but changes with $n$.
In particular, in the increasing field the distance between consecutive jumps decreases, whereas in the decreasing field, on the contrary, the distance between jumps
increases. This conclusion can be compared with the experimental results shown in Figs.\
\ref{fig3} and \ref{fig5}. For this purpose we plot the inverse of $\Delta \phi_n^{\pm}$ as a
function of $n$ in Fig.\ \ref{theory_fig2} \cite{plouf}. Since $1/\Delta \phi_n^{\pm}$ roughly corresponds
to the position of the maximum of the Fourier transform of $C_p(H)$, the red solid line and the blue dashed line of Fig. \ref{theory_fig2} can be compared to Figs.\ \ref{fig3} and \ref{fig5} respectively. 
The predicted variation of the periodicity of heat capacity variations appears to be of the
order of 10\%, similarly to the experimentally observed values.
Note that according to Eq.\ (\ref{dphipm}),
the magnitude of this effect is controlled by the ratio $w/R$ and that the effect disappears
and the jumps of $C_p(\phi)$ become equally spaced in the limit $w \rightarrow 0$.

In the above analysis we put $k = n \pm 1$, although nothing justifies such a choice.
In general, solving $A = 0$ for $\phi$ with arbitrary $k$ yields a (rather lengthy) expression for the critical magnetic fluxes at which the
$n$-vortex state becomes unstable with respect to an admixture of the $k$-vortex state.
In the limit $w/R \rightarrow 0$ these are

\begin{eqnarray}
\phi_{nk}^{\pm} = 2 n - k \pm \sqrt{2(n-k)^2 + [R/\xi(T)]^2}
\label{phipmk}
\end{eqnarray}

As the magnetic field modulus increases (decreases) we expect a transition to a state with $k > n$ ($k < n$), and hence the `$+$' (`$-$') sign should be chosen in Eq.\ (\ref{phipmk}).
The state $k$ with respect to which the instability occurs first can be estimated by searching for a minimum
(maximum) 
of $\phi_{nk}^{\pm}$ as a function of $k$. This yields $k = n \pm R/\sqrt{2} \xi(T)$. We see therefore that at temperatures
$T$ close to $T_c$, when $\xi(T) \rightarrow \infty$ and $R/\xi(T) \lesssim 1$, transitions to the states
$k = n \pm 1$ should indeed be favored. In contrast, at lower temperatures, when $\xi(T)$ becomes smaller than $R$,
transitions to states $k = n \pm 2$, $n \pm 3$, etc. can occur. Such `multivortex jumps' have indeed been reported
in Ref.\ \cite{bourgprl} and earlier in magnetic measurements of other groups \cite{pedersen,vodo03}. They have also been
analyzed theoretically \cite{vodo02}. This reminds us once again that these multivortex jumps or `flux avalanches' are related to the existence of
the metastable states and to the possibility for the system to remain in these states for a sufficiently long time. 

Another important comment is connected with the stability analysis presented above.
As first noted by Horane \textit{et al.} \cite{horane96}, the stability analysis of a given $n$-vortex
state $\psi_n(\vec{r})$ of a superconducting ring should not, in general, rely on a particular functional form
[$\psi_k(\vec{r})$ in our case] of the perturbation. By fixing the latter, we unavoidably overestimate the
stability of the state under consideration. Unfortunately, the full treatment of the stability problem appears to be
mathematically too involved and we therefore have chosen to present a less general but much more physically transparent analysis here. Such an analysis still yield qualitatively correct conclusions.

\section{Discussion}

After the exposition of the experimental results, the data treatment and the development of the theoretical model describing our system, a deeper discussion of the main results is needed. Through the calculation of the SPWVD of the heat capacity signal versus magnetic field, it has been shown that the $C_p(H)$ oscillations are not stationary. The periodicity of the jumps strongly depends on the sweeping direction of the magnetic field (increasing or decreasing). From Fig. 3 and Fig. 5, it is clear that when magnetic vortices are expelled from the loops (by decreasing the absolute value of the magnetic field), the heat capacity jumps are far from each other, giving a low-frequency oscillation of $C_p(H)$. On the other hand, when the magnetic field is increased, vortices penetrate the loops and the jumps are closer and closer to each other, giving high-frequency heat capacity oscillations. This non-stationarity of the heat capacity oscillations is not negligible because from the experimental data, as well as from the theoretical estimation, the variation of the periodicity can be greater than 10\% between expulsion of the first vortex and the penetration of the last one.

In other words, and restraining our discussion only to positive magnetic fields, it can be said that as the field is increased, the system tends to spend less and less time in each successive giant vortex state. Furthermore when the vorticity $n$ becomes high it appears from  Fig. \ref{theory_fig1} that the occupied states are the stable ones, in contrast to the low field regime where the system almost always evolves along metastable states. This means that the higher the vorticity $n$, the easier it is to add another vortex into the loop, until the critical field is reached and the superconductivity suppressed. Thus the energy barriers the systems has to overcome to jump from the $n$ to the $n+1$ state tend to disappear at higher fields. In that last case, the contour through which the fluxoid is quantized is close to the external edge of the loop and the supercurrents are localized near the outer boundary of the superconductor, leading to the "traditional" picture of the giant vortex state regarded as a surface superconductivity state \cite{saintjames63}.

On the other hand, in decreasing field, when the first vortices are expelled out of the loops (high magnetic field), the successive jumps occur with a large period, meaning that metastable states can survive for a long time before the energy barrier between the $n$ and $n-1$ states is suppressed. This feature is confirmed by Fig. \ref{theory_fig1}: near the critical field the system spends more time in metastable states when vortices are expelled than it does in increasing field. In this situation the apparent contour through which the fluxoid is quantized is close to the inner boundary of the loop; the kinetic energy of the supercurrent carriers is higher than it would be if the system was in its ground state, where the supercurrents are localized at the outer boundary.

Furthermore the differences between penetration and expulsion of vortices are enhanced due to the fact that in increasing field the surface barrier can be destroyed by surface defects, which is not the case in decreasing field \cite{BeanLivingston}. This effect has not been taken into account in the theoritical work leading to Eq. \ref{dphipm}, but it is in good agreement with the small deviation of the experimental data presented in Fig. \ref{fig3} from the linear behavior predicted by Eq. \ref{dphipm} and Fig. \ref{theory_fig2}.

\section{Conclusions} 

We have studied the heat capacity behavior of an assembly of non-interacting superconducting loops in applied magnetic field. Heat capacity discontinuities are observed when the magnetic field is swept. The jumps are associated to vortex expulsion or vortex entrance in the loops. We have studied the variation of the periodicity of the heat capacity jumps versus the magnetic field. Using the signal processing tool based on Wigner-Ville distribution, we were able to image the variation of that periodicity versus the direction of the magnetic field sweep. As the field is increased, the vortices entering the loops are not submitted to the same barrier of energy as when the field is decreased and the vortices are expelled. The GL theory describes our experimental observations with reasonable accuracy.

\section*{Acknowledgments}

We would like to thank E. Andr\'e, P. Lachkar, J-L. Garden, C. Lemonias, B. Fernandez, T. Crozes for technical support, P. Brosse-Maron, T. Fournier, Ph. Gandit and J. Richard for fruitfull discussions and help. We thank the Région Rhône-Alpes and the IPMC (Institut de Physique de la Mati\`ere Condens\'ee) of Grenoble for financing a part of this project.


\newpage

\begin{thebibliography}{99}


\bibitem{riou97}
O. Riou, P. Gandit, M. Charalambous, and J. Chaussy,
\textit{Rev. Sci. Instrum.} \textbf{68,} 1501 (1997).

\bibitem{fom97}
F. Fominaya, T. Fournier, P. Gandit, and J. Chaussy,
\textit{Rev. Sci. Instrum.} \textbf{68,} 4191 (1997).

\bibitem{lindell}
A. Lindell, J. Mattila, P.S. Deo, M. Manninen and J. Pekola
\textit{Physica B} \textbf{284,} 1884 (2000).

\bibitem{bourgprl}
O. Bourgeois, S.E. Skipetrov, F. Ong, and J. Chaussy
\textit{Phys. Rev. Lett.} \textbf{94,} 057007 (2005).

\bibitem{ongPRB2006}
F.R. Ong, O. Bourgeois, S.E. Skipetrov, and J. Chaussy
\textit{Phys. Rev. B} \textbf{74,} 140503(R) (2006).

\bibitem{rouknano}
W. Chung Fon, K.C. Schwab, J.M. Worlock, and M.L. Roukes
\textit{Nano Lett.}
\textbf{5,} 1968 (2005).

\bibitem{buzdin95}
A. Bezryadin, A. Buzdin, and B. Pannetier,
\textit{Phys. Rev. B} \textbf{51,} 3718 (1995).

\bibitem{fink81}
H.J. Fink and V. Gr\"{u}nfeld
\textit{Phys. Rev. B} \textbf{23,} 1469 (1981).

\bibitem{buzdin2005}
A. Buzdin 
\textit{Phys. Rev. B} \textbf{72,} 100501(R) (2005).

\bibitem{valls2005}
P.H. Barsic, O.T. Valls and K. Halterman
\textit{Cond-Mat} 0512285 (2005).

\bibitem{vodo03}
D.Y. Vodolazov, F.M. Peeters, S.V. Dubonos, and A.K. Geim,
\textit{Phys. Rev. B} \textbf{67,} 054506 (2003).

\bibitem{baelus01}
B.J. Baelus, F.M. Peeters, and V.A. Schweigert,
\textit{Phys. Rev. B} \textbf{63,} 144517 (2001).

\bibitem{mosh}
V.V. Moshchalkov, L. Gielen, C. Strunk, R. Jonckheere, X. Qiu, C. Van Haesendonck and Y. Bruynseraede
\textit{Nature} \textbf{373}, 319 (1995).


\bibitem{morelle}
M. Morelle, V. Bruyndoncx, R. Jonckheere, and V.V. Moshchalkov
\textit{Phys. Rev. B} \textbf{64,} 064516 (2001).

\bibitem{zhang97}
X. Zhang and J.C. Price,
\textit{Phys. Rev. B} \textbf{55,} 3128 (1997).

\bibitem{pannet}
B. Pannetier, A. Bezryadin, and A. Eichenberger, \textit{Physica B} \textbf{222,} 253 (1996).

\bibitem{pedersen}
S. Pedersen, G.R. Kofod, J.C. Hollingbery, C.B. Sorensen, and P.E. Lindelof,
\textit{Phys. Rev. B} \textbf{64,} 104522 (2001).

\bibitem{geim}
A.K. Geim, I.V. Grigorieva, S.V. Dubonos, J.G.S. Lok, J.C. Maan, A.E. Filippov, and F.M. Peeters,
\textit{Nature} \textbf{390,} 259 (1997).

\bibitem{gandit}
P. Gandit, J. Chaussy, B. Pannetier, A. Vareille, and A. Tissier
\textit{EuroPhys. Lett.} \textbf{3,} 623 (1987).

\bibitem{buisson}
O. Buisson, P. Gandit, R. Rammal, Y.Y. Wang, and B. Pannetier
\textit{Phys. Lett. A} \textbf{150,} 36 (1990).

\bibitem{deo}
P. Singha deo, J.P. Pekola, and M. Manninen
\textit{EuroPhys. Lett.} \textbf{5,} 649 (2000).

\bibitem{sullivan}
P.F. Sullivan and G. Seidel
\textit{Phys. Rev.} \textbf{173,} 679 (1968).


\bibitem{baelus05}
B.J. Baelus, A. Kanda, F.M. Peeters, Y. Ootuka, and K. Kadowaki
\textit{Phys. Rev. B} \textbf{71,} 140502(R) (2005).

\bibitem{signproces}
W.J. Mecklenbrauker  and F. Hlawatsch , \textit{The Wigner Distribution} (Elsevier eds, Amsterdam, 1997).

\bibitem{harris}
F.J. Harris,
\textit{Proc. of I.E.E.E.} \textbf{66,} 51 (1978).

\bibitem{tink04}
M. Tinkham,
\textit{Introduction to Superconductivity} (2nd edition, Dover, 2004).

\bibitem{fomin98}
V.M. Fomin, V.R. Misko, J.T. Devreese, V.V. Moshchalkov,
\textit{Phys. Rev. B} \textbf{58,} 11703 (1998).

\bibitem{vodo02}
D.Y. Vodolazov and F.M. Peeters,
\textit{Phys. Rev. B} \textbf{66,} 054537 (2002).

\bibitem{horane96}
E.M. Horane, J.I. Castro, G.C. Buscaglia, and A. Lopez,
\textit{Phys. Rev. B} \textbf{53,} 9296 (1996);
J.I. Castro and A. Lopez,
\textit{Phys. Rev. B} \textbf{72,} 224507 (2005).

\bibitem{plouf}
We calculate $1/\Delta \phi_n^\pm$ up to second order in $(w/2R)$. This yields the same result as Eq. \ref{fluxoid} but with an opposite sign in front of the second term.

\bibitem{saintjames63}
D. Saint-James and P.-G. de Gennes,
\textit{Phys. Lett.} \textbf{7,} 306 (1963).

\bibitem{BeanLivingston}
C.P. Bean and J.B. Livingston,
\textit{Phys. Rev. Lett.} \textbf{12,} 14 (1964).

\end{thebibliography}
\end{document}